\documentclass[journal=ancac3,manuscript=article]{achemso}

\usepackage[T1]{fontenc} 
\usepackage{amsmath,amssymb} 
\usepackage{soul}               
\usepackage{color,colordvi}     
\usepackage{makecell}
\usepackage{pdfpages}
\mciteErrorOnUnknownfalse

\setkeys{acs}{
  abbreviations = true,
  biblabel      = fullstop,
  etalmode      = truncate,
  maxauthors    = 20,
  keywords      = true
}

\author{Fabrizio Camerin}
\affiliation{CNR Institute for Complex Systems, Uos Sapienza, Piazzale Aldo Moro 2, 00185 Roma, Italy}
\alsoaffiliation{Department of Basic and Applied Sciences for Engineering, Sapienza University of Rome, via Antonio Scarpa 14, 00161 Roma, Italy}
\email{fabrizio.camerin@uniroma1.it}

\author{Miguel {\'A}ngel Fern{\'a}ndez-Rodr{\'\i}guez}
\affiliation{Laboratory for Interfaces, Soft Matter and Assembly, Department of Materials, ETH Z{\"u}rich, Vladimir-Prelog-Weg 5, 8093 Z{\"u}rich, Switzerland}

\author{Lorenzo Rovigatti}
\affiliation{Department of Physics, Sapienza University of Rome, Piazzale Aldo Moro 2, 00185 Roma, Italy}
\alsoaffiliation{CNR Institute for Complex Systems, Uos Sapienza, Piazzale Aldo Moro 2, 00185 Roma, Italy}

\author{Maria-Nefeli Antonopoulou}
\affiliation{Laboratory for Interfaces, Soft Matter and Assembly, Department of Materials, ETH Z{\"u}rich, Vladimir-Prelog-Weg 5, 8093 Z{\"u}rich, Switzerland}

\author{Nicoletta Gnan}
\affiliation{CNR Institute for Complex Systems, Uos Sapienza, Piazzale Aldo Moro 2, 00185 Roma, Italy}
\alsoaffiliation{Department of Physics, Sapienza University of Rome, Piazzale Aldo Moro 2, 00185 Roma, Italy}

\author{Andrea Ninarello}
\affiliation{CNR Institute for Complex Systems, Uos Sapienza, Piazzale Aldo Moro 2, 00185 Roma, Italy}
\alsoaffiliation{Department of Physics, Sapienza University of Rome, Piazzale Aldo Moro 2, 00185 Roma, Italy}

\author{Lucio Isa}
\affiliation{Laboratory for Interfaces, Soft Matter and Assembly, Department of Materials, ETH Z{\"u}rich, Vladimir-Prelog-Weg 5, 8093 Z{\"u}rich, Switzerland}
\email{lucio.isa@mat.ethz.ch}

\author{Emanuela Zaccarelli}
\affiliation{CNR Institute for Complex Systems, Uos Sapienza, Piazzale Aldo Moro 2, 00185 Roma, Italy}
\alsoaffiliation{Department of Physics, Sapienza University of Rome, Piazzale Aldo Moro 2, 00185 Roma, Italy}
\email{emanuela.zaccarelli@cnr.it}

\title{Microgels Adsorbed at Liquid-Liquid Interfaces:
A Joint Numerical and Experimental Study}


\begin{document}

\begin{abstract}
\noindent
Soft particles display highly versatile properties with respect to hard colloids, even more so at fluid-fluid interfaces. In particular, microgels, consisting of a cross-linked polymer network, are able to deform and flatten upon adsorption at the interface due to the balance between surface tension and internal elasticity. Despite the existence of experimental results, a detailed theoretical understanding of this phenomenon is still lacking due to the absence of appropriate microscopic models. In this work, we propose an advanced modelling of microgels at a flat water/oil interface. The model builds on a realistic description of the internal polymeric architecture and single-particle properties of the microgel and is able to reproduce its experimentally observed shape at the interface.
Complementing molecular dynamics simulations with \textit{in-situ} cryo-electron microscopy experiments and atomic force microscopy imaging after Langmuir-Blodgett deposition, we compare the morphology of the microgels for different values of the cross-linking ratios. Our model allows for a systematic microscopic investigation of soft particles at fluid interfaces, which is essential to develop predictive power for the use of microgels in a broad range of applications, including the stabilization of smart emulsions and the versatile patterning of surfaces.
\bigskip
\begin{figure}[h!]
\centering
\includegraphics[scale=1.2]{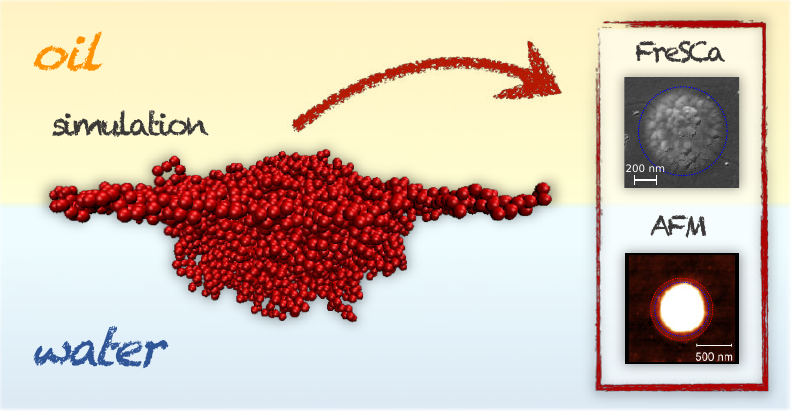}
\end{figure}

\end{abstract}


\bigskip
\bigskip
One of the main goals of soft matter science is to take advantage of the microscopic complexity of nanoscale and colloidal building blocks  in order to design the macroscopic properties of an emerging material. Within the class of soft colloids, those possessing an intrinsic polymeric structure are among the best candidates to exploit this connection, thanks to their deformability, elasticity and possibility of interpenetration~\cite{vlassopoulos2014tunable}. Altogether, these features enable soft particles to display a much richer behavior as compared to their hard counterparts\cite{vlassopoulos2014tunable, lyon2012polymer}.  Besides the structural aspects, one of the main advantages of soft polymeric particles is their ability to respond to environmental stimuli. This capability has not only attracted significant attention for fundamental studies\cite{likos2006soft,gnan2018microscopic}, but has also triggered several chemical, medical and biological applications\cite{liu2006deformation,rivest2007microscale, liu2016methods,doring2013responsive,foster2017protease}. Microgels, on which we focus hereinafter, represent one of the most intriguing choices in this regard.

Microgels are cross-linked polymer networks whose properties depend on the nature of the constituent monomers. One of the most studied cases is that in which polymers are thermo-responsive, usually based on Poly-N-isopropylacrylamide (PNIPAM), resulting in particles that display a so-called Volume Phase Transition (VPT), from a low-temperature swollen state to a high-temperature collapsed state\cite{fernandez2011microgel, saunders1999microgel}.

\begin{figure}[t!]
\centering
\includegraphics[scale=0.62]{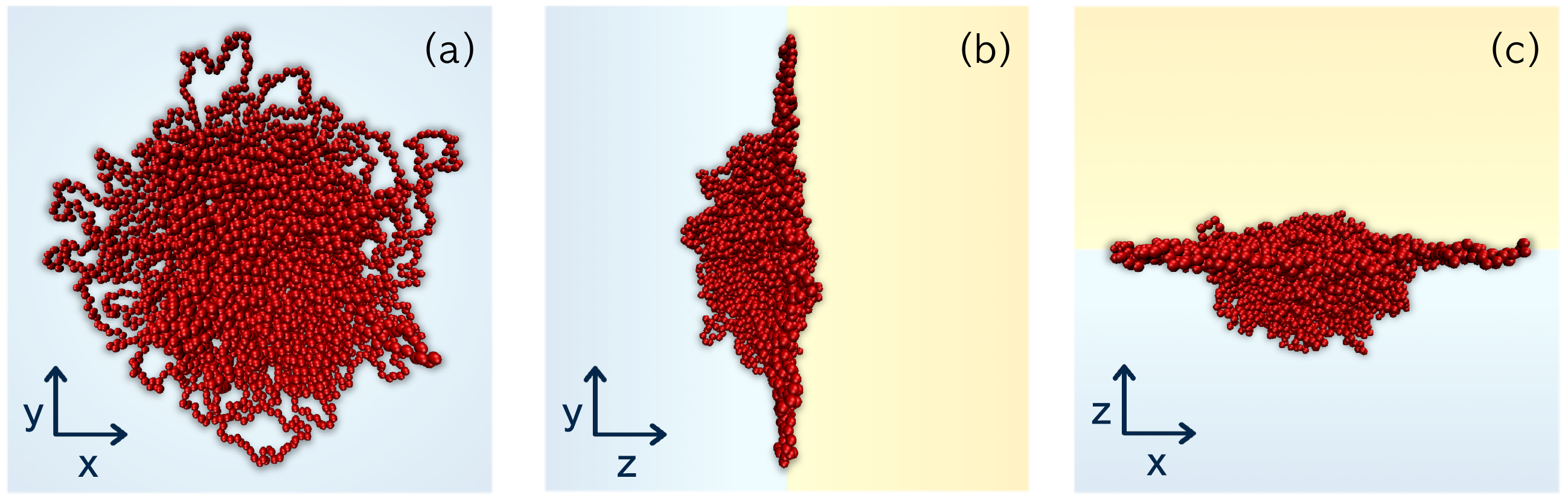}
\caption{\small \textbf{Simulation snapshots of a microgel at a water/oil interface} in the three different planes of observation: (a) top view (interface plane $xy$), (b,c) side views in which $z<0$  corresponds to the water region and $z>0$ to the oil one, respectively. The observed conformation is loosely called ``fried-egg'' shape.}
\label{fig:snap5k}
\end{figure}

The potentialities of microgels are not limited to bulk applications, and a promising research field is opening up to exploit these particles at the interface between two immiscible liquids. In general, fluid interfaces constitute ideal settings where nanocomposites and colloidal particles can accumulate and self-assemble\cite{niu2010synthesis,binks2017colloidal,isa2017two,ballard2018colloidal,qin2018structure,mehrabian2016soft}. Recently, microgels have also been experimentally explored under these conditions,\cite{fujii2005stimulus,li2013microgel,rey2016isostructural,rey2017interfacial,schulte2018probing} where the colloid-polymer duality of such particles\cite{lyon2012polymer} strongly manifests. Indeed, their internal polymeric structure allows them to spread and flatten at the interface in order to maximize their area, reduce non-favourable contacts between the two liquids and thus lower their surface tension\cite{richtering2012responsive}. This phenomenology is always dictated by the elasticity of the single objects, in contrast with hard particles where the latter does not play a role. Interestingly, experimental images of microgels at water/oil interfaces\cite{geisel2012unraveling,style2015adsorption,kwok2016confocal,geisel2014new} have evidenced a preferential protrusion of the particle center on the water side. This feature is the result of two main contributions: first of all, the higher solubility of PNIPAM chains in water makes the microgel to maximize the surface exposed to water; secondly, the fact that the cross-linking density of the microgel is usually not homogeneous and decreases towards the periphery of the particle\cite{aufderhorst2018nanoscale,conley2017jamming} implies that the inner core mostly retains its spherical shape also at the interface. As a result, the peculiar conformation of microgels at the liquid-liquid interface is usually referred to as a ``fried-egg'' shape\cite{wellert2015responsive}, as shown in Figure~\ref{fig:snap5k}.

The combination of responsiveness and flexibility leads to several advantages for applications at interfaces. Contrarily to traditional Pickering emulsions\cite{richtering2012responsive}, where rigid particles adsorb at the interface\cite{chevalier2013emulsions}, the use of microgels would provide temperature -- or pH -- sensitive emulsions that could be stabilized or destabilized on demand by directly changing these control parameters\cite{schmidt2011influence,liu2012non}. Moreover, microgels' deformability may be exploited for nanostructuring elements or for other high-end applications such as sensing, interferometry and biocatalysis\cite{wiese2013microgel,dickinson2015microgels,rey2015fully,kim2005bioresponsive,tsuji2005colored}. 

However, a detailed microscopic description of the conformation of microgels at fluid interfaces is still missing. This is due to the fact that numerical and theoretical studies of microgels in bulk have been limited for a long time to unrealistic models, such as the coarse-grained description provided by the Hertzian model\cite{ghosh2018linear,mohanty2014effective} and the monomer-resolved diamond network\cite{kobayashi2017polymer,ghavami2016internal,keidel2018time,brugnoni2018swelling}.
Only recently, a few realistic {\it in silico} models of microgels have been proposed\cite{gnan2017silico,moreno2018computational,rovigatti2019numerical}. In particular, some of us have developed a model, based on a disordered network, aimed at reproducing in detail the single-particle properties and the swelling behavior\cite{gnan2017silico} for different cross-linker fractions $c$ and internal topologies. In a recent extension of the model\cite{ninarello2019advanced}, a fine control on the microgel internal density profiles was obtained by adding a designing force on the cross-linkers which drives them towards the core as in the experimental synthesis, making it possible to tune the core-corona relative extent in closer agreement to experiments, independently of the microgel size.

Building on this
model, so far only studied in bulk, here we provide a comprehensive modelling of a single PNIPAM microgel particle at a flat water/oil interface. This is an experimentally relevant condition that constitutes the precursor of a water/oil emulsion and also captures some of the salient features of processes that exploit self-assembly and deposition from macroscopically flat fluid interfaces, \textit{e.g.} Langmuir-Blodgett processes~\cite{petty1996langmuir}.  The model explicitly includes the two solvents and quantitatively accounts for the surface tension between them. Numerical results are directly compared with experiments, where microgels are imaged \textit{in-situ} at the water/oil interface using a cryo-scanning electron microscopy (SEM) or are inspected after deposition from the interface onto a silicon wafer by means of Atomic Force Microscopy (AFM). A good agreement between experiments and simulations is found for different cross-linker concentrations, which makes it possible to carefully assess the role played by the stiffness on the microgel structure at the interface. We first show results for the water/hexane interface, demonstrating that the explicit-solvent microgel model developed in this work is able to capture the physical details of single soft particles adsorbed at a flat interface. To provide robustness to our approach, we also perform additional simulations and experiments at the water/benzene interface, which has a significantly lower surface tension, again finding good agreement between the two. Interestingly, we find that the spreading of the microgel remains mostly unaltered for both conditions, a result which provides further physical insights about the adsorption mechanism of polymer-based objects.

\section{Results and discussion}

In this section we present the main results of this work. First of all, we discuss the parameterization of the model.
This first step is a fundamental procedure that requires a careful understanding of the role of the solvent-solvent and monomer-solvent interactions and should always be carried out by closely comparing with experimental results. Next, we report the detailed investigation of the size and shape variations of the microgel as a function of cross-linker concentration at the water/hexane interface for both   simulation and experiments. Doing so, we also discuss the differences between the present study based on a disordered realistic microgel and those that employ a regular diamond network. Finally, we extend our approach to a different interface, namely water/benzene, that displays a lower surface tension.

\subsection{Modelling the microgel-solvent interactions}
\label{sec:dependence}

We start by analyzing in detail the choice of the microgel-solvent interaction parameters in simulations and their interplay with the surface tension between the two solvents. The latter is fixed in order to reproduce a realistic surface tension, as detailed in the Methods section.

The main contributions to the free energy that dictate the shape of a PNIPAM microgel adsorbed at an interface are~\cite{geisel2012unraveling}: (i) the tendency to maximise its surface so as to minimise the solvent-solvent interface; (ii) the higher affinity for water with respect to oil, which makes the polymer chains to organize in such a way to be mostly solvated by water; (iii) the elasticity of the microgel, which acts against changes in volume and shape. Given the disordered and inhomogeneous nature of the microgels, the interplay between these three contributions is non-trivial and hard to quantify \textit{a priori}, although there exist theoretical models that can help in detecting qualitative trends~\cite{rumyantsev2016polymer,vasudevan2018dynamics}.
\begin{figure}[b!]
\centering
\includegraphics[scale=0.72]{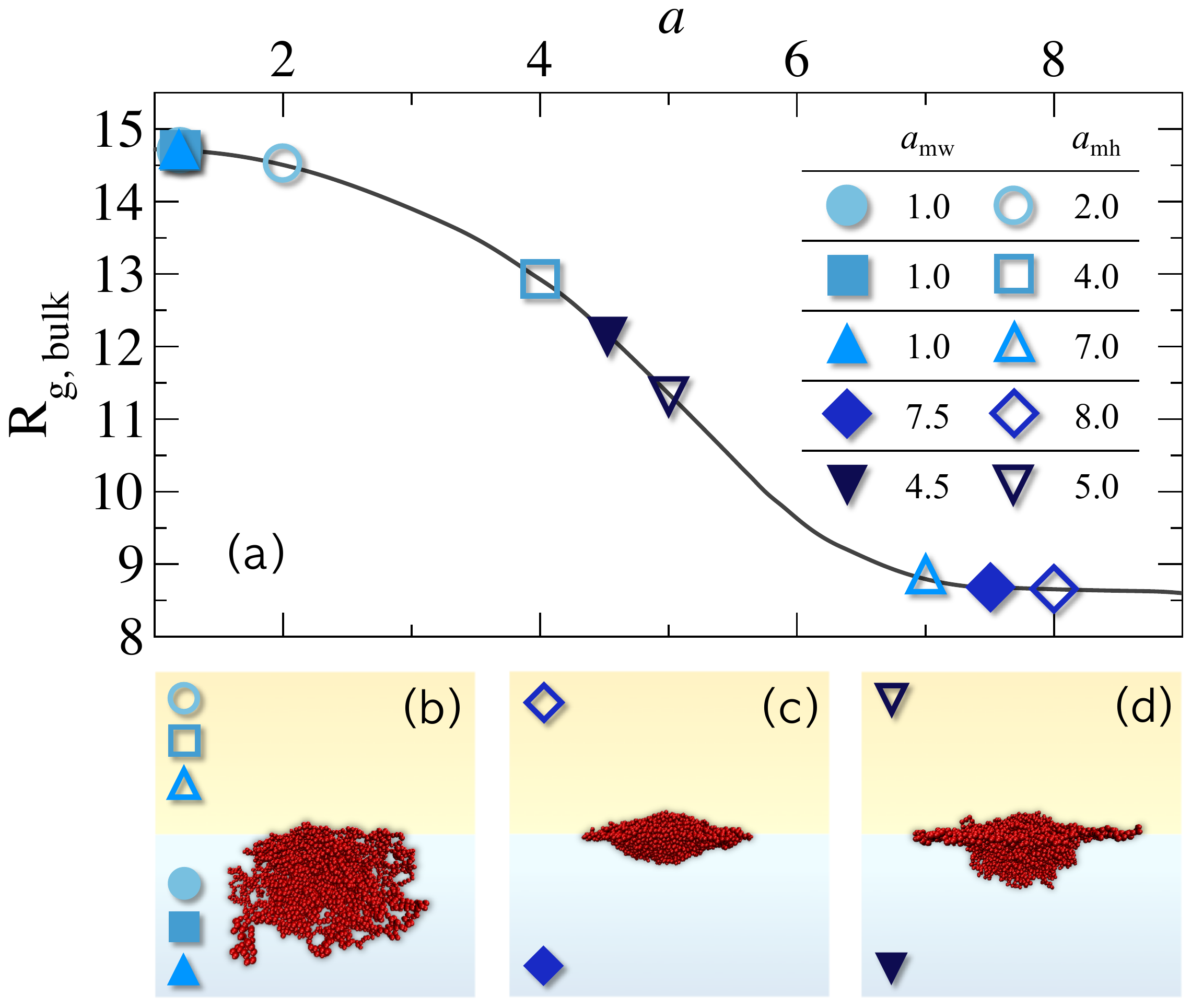}
\caption{\small \textbf{Choice of the monomer-solvent interaction parameters.} (a) Radius of gyration $R_{g, bulk}$ of a microgel with  cross-linker concentration $c = 3.8\%$ in a one-component bulk fluid with solvophobic parameter $a$. Along with the calculated swelling curve (black solid line), pairs of solvophobic parameters for monomer-water ($a_{\rm mw}$, full symbols) and monomer-hexane ($a_{\rm mh}$, empty symbols) interactions, which we analyze in interfacial simulations, are highlighted and listed in the inset table. As expected, for the three choices where $a_{\rm mw}=1.0$, $R_{g, bulk}$ coincides (the corresponding filled symbols are superimposed onto each other). The maximum extension of the microgel on the plane of the interface is obtained for the combination of parameters $a_{\rm mw}=4.5, a_{\rm mh}=5.0$. Representative snapshots (b), (c), (d) are side views that exemplify the conformation assumed by the microgel at the interface for different $a_{\rm mw}$ -- $a_{\rm mh}$ choices.}
\label{fig:swelling}
\end{figure}

We use a Dissipative Particle Dynamics (DPD) description, as detailed in Methods, which employs two solvophobic parameters to control the affinity of the monomers towards each of the two solvents. These are $a_{\rm mw}$ (monomer-water) and $a_{\rm mh}$ (monomer-hexane): the higher $a_{\rm mw}$ and $a_{\rm mh}$, the more solvophobic the polymers are with respect to the corresponding solvent particles. To choose appropriately the values of the solvophobic parameters, we first need to calculate the swelling curve of a single microgel in a one-component bulk fluid where the monomer-solvent interaction is controlled by a single solvophobic parameter $a$, which will be later used to mimic monomer-water and monomer-hexane interactions, respectively.
The corresponding radius of gyration $R_{g,bulk}$ for a microgel with cross-linker concentration $c = 3.8\%$ is shown in Figure~\ref{fig:swelling}(a) as a function of $a$, quantifying the size variation of the microgel with the change in solvent affinity. \textit{Via} this procedure, we determine the range of $a$-values for which the microgel goes from a maximally swollen case ($a \approx 1$, below which the coupling between the solvent and the monomers would be too small to properly thermalize the microgel) to a state where the majority of the solvent is expelled from the network and the microgel has collapsed ($a \approx 8$). This is the range within which $a_{\rm mw}$ and $a_{\rm mh}$ should be chosen. Considering the higher affinity for water, it is clear that one should set $a_{\rm mw} < a_{\rm mh}$, as these two values directly control the effective monomer-water and monomer-hexane interactions. However, the balance between $a_{\rm mw}$ and $a_{\rm mh}$ and their interplay with the given water/oil surface tension produce nontrivial effects, as shown in the following.

\noindent First of all, it is important to note that for $a_{\rm mw} \sim 1$ the microgel never takes the ``fried-egg'' shape.  
In particular, we test three different combinations that comprise $a_{\rm mw}=1$ and a value of $a_{\rm mh} > a_{\rm mw}$, as indicated schematically in Fig.~\ref{fig:swelling}(a). Under all these conditions (see Figure~\ref{fig:swelling}(b)) the microgel only partially adsorbs at the interface, remaining mostly in the water region and retaining a quasi-spherical shape. This is explained by a too small free energy gain provided by the spreading of the particle and the reduction of the water/oil contact surface with respect to the elastic and entropic contributions of the microgel, that are consequently found to dominate the microgel behavior under these conditions. By contrast, choosing high values of both $a_{\rm mh}$ and $a_{\rm mw}$, the bad quality of two solvents makes the microgel collapse onto itself, taking a lens-shaped conformation, as shown in Figure~\ref{fig:swelling}(c). In this case, the microgel interacts in a rather similar manner with both solvents and a difference in protrusion on the water side, despite being present, is barely noticeable.

\noindent It is only for intermediate values of the solvophobic parameters that the elastic free energy contribution can be overcome by the interfacial term, which is strong enough to make the microgel spread over the interface to minimise the contact surface between the two solvents. In addition, the not-so-high solvophobicity now also allows the microgel to present a clear preference for water with respect to oil, thus giving rise to a well-defined core-centered protrusion in the water phase and a nearly zero protrusion into the oil. Figure~\ref{fig:swelling}(d) shows a simulation snapshot of a microgel taking the ``fried-egg'' conformation obtained by choosing $a_{\rm mw} = 4.5$ and $a_{\rm mh} = 5.0$, which are the values we will use for the microgel-water/hexane system throughout the next section.

\subsection{Characterization of the microgels at the liquid-liquid interface}
The typical microstructure of an interfacial microgel, resulting from the interplay between 
particle architecture and surface tension, is reproduced in Fig.~\ref{fig:snap5k}. We recall that the internal structure of microgels comprises of a rather homogeneous core, with a higher density of cross-linkers, and a loose corona complemented by a non-negligible number of dangling chains, where the number of cross-linkers is rather low\cite{aufderhorst2018nanoscale,conley2017jamming}.  As a consequence, a clear flattening of the corona takes place at the interface, which exposes the core, giving rise to a protrusion in the center of the microgel. 
A realistic modeling of the internal degrees of freedom appears to be crucial to reproduce such phenomenology. Indeed, it is the polymeric, inhomogeneous nature of the system that allows microgels to deform and assume the ``fried-egg'' shape. 
\begin{figure}[b!]
\centering
\includegraphics[scale=0.54]{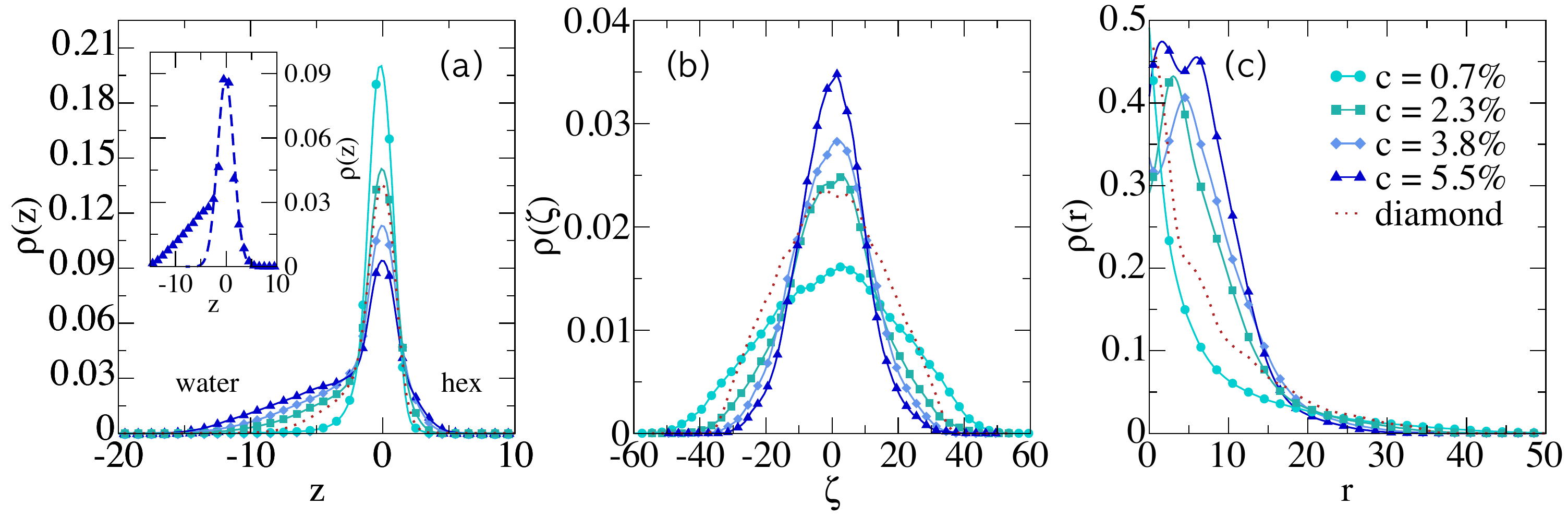}
\caption{\small \textbf{Density profiles at the interface} for disordered microgels with $c=0.7, 2.3, 3.8, 5.5\%$ and for the diamond-lattice-based microgel with $c=5\%$. Panel (a) shows the density profiles $\rho(z)$ obtained by binning the simulation box parallel to the interface.  The inset shows a gaussian fit (dashed line) of the portion of microgel that stands at the interface for $c=5.5\%$. Panel (b) reports $\rho(\zeta)$ and shows how the microgel flattens at the interface. Panel (c) shows the radial density profiles $\rho(r)$ taken with respect to the center of mass of the microgel; the same four cross-linker ratios are analyzed. Lines are guides to the eye.  All data are normalized to the average number of particles of the $c=5.5\%$ microgels.}
\label{fig:densprof}
\end{figure}
\\
More information on how the microgel arranges itself at the interface is gained by looking at the density profiles reported in Figure~\ref{fig:densprof}.
In particular, Figure~\ref{fig:densprof}(a) displays the $\rho(z)$ density profile, calculated at a distance $z$ from the interface, and obtained by dividing the simulation box along the $z$-axes into three-dimensional bins that are parallel to the interface. Figure~\ref{fig:densprof}(b) shows instead the $\rho(\zeta)$ density profile, where bins are taken orthogonally to the interfacial $xy$-plane and $\zeta=x, y$. This is calculated at distance $\zeta$ with respect to the center of mass of the microgel and averaged over the two directions. Finally, the radial density profiles $\rho(r)$, obtained by building spherical shells at distance $r$ from the center of mass of the microgels, are reported in Figure~\ref{fig:densprof}(c), providing information on the core size of the microgels. In each panel, we report the density profiles for four different values of the cross-linker ratio $c$, namely $0.7, 2.3, 3.8$ and $5.5\%$. For completeness, we also provide results for a diamond network microgel with a representative cross-linker concentration $c=5$\%.
\\
We start by discussing $\rho(z)$. Two main regions can be recognized, depending on the value of $z$. The first one is the central part of the profile, which corresponds to the section of the microgel that builds up at the interface. As expected, the maximum density is found at $z=0$ owing to the greater number of monomers that are present at the interface.
The extent of this part, which determines the interfacial region, can be properly identified by a gaussian fit to the data, shown in the inset of Fig.~\ref{fig:densprof}(a) for $c=5.5\%$, which captures all the signal on the oil side for all studied values of $c$ and confirms the poor solubility of the polymeric material in oil. The second one, for more negative $z$, comprises the protrusion of the microgel in water, which strongly depends on the cross-linker concentration. Indeed, the more the microgel is cross-linked, the more its core is pronounced, giving rise to an increasingly asymmetric tail in the profiles on the water side.  
\\
Looking at $\rho(\zeta)$ instead, the highest density is found at the center of the interface ($\zeta=0$) due to the presence of the core.  The distributions become broader and broader as $c$ decreases, since the difference between the core and corona regions is less defined when the number of cross-linkers is small. The same features are also confirmed by the radial profiles $\rho(r)$, where a stronger initial bump signals the presence of a denser, well-defined core, which is indicative of a ``fried-egg'' shaped microgel. For $c<1\%$ this feature is found to be almost absent, while it manifestly develops for $c\geq2.3\%$. 
\\
Comparing with the regular diamond network~\cite{mourran2016colloidal,rumyantsev2016polymer,gumerov2016mixing} with $c=5\%$, for which the density profiles are also reported in Fig.~\ref{fig:densprof}, we find that even for this high value of cross-linker concentration a well-defined core is not present. This is also clearly visible in the snapshots reported in the SI.  Interestingly, the microgel extension at the interface is even larger than that of the disordered one, again due to the absence of the core. These effects produce an unrealistic conformation of the diamond microgel at the interface, which resembles the one assumed by the disordered network with a much lower cross-linking ratio. Thus, for the regular topology, the ``fried-egg'' shape is not observed in simulations, limiting the applicability of the diamond model in the description of real particles at interfaces.
\\
Experimentally, we have access to indirect measurements of the lateral microgel size at the water/hexane interface after deposition onto a solid substrate (silicon wafer) \cite{geisel2012unraveling}. In this way, the microgels dry out but, following previous works~\cite{geisel2014highly,rey2016isostructural}, we can assume that they retain the same extension they had at the interface also under these conditions. The height profiles of dried microgels are reported in Fig.~\ref{fig:height} as a function of the cross-linking ratios. Qualitatively, they show the same behavior as observed in $\rho(\zeta)$ calculated with simulations (Fig.~\ref{fig:densprof}), presenting a lower degree of spreading as the cross-linking ratio increases.
\begin{figure}[t!]
\centering
\includegraphics[scale=0.35]{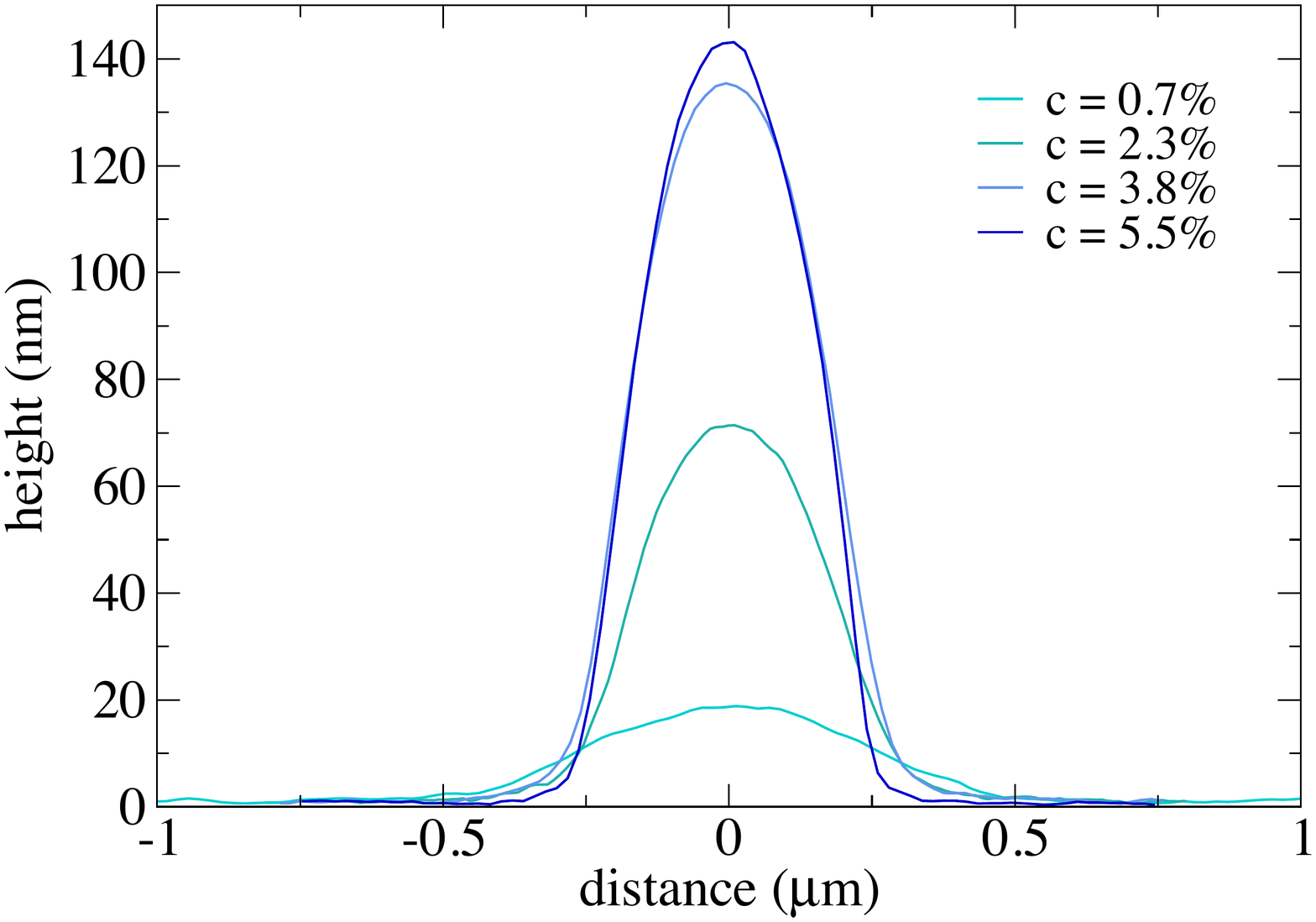}
\caption{\small \textbf{AFM height profiles of dry isolated microgels} deposited onto silicon wafers, extracted from AFM images, for different values of the cross-linker ratios $c$.}
\label{fig:height}
\end{figure}
\\
In order to assess the validity of the theoretical model, we perform a qualitative comparison with experiments, in terms of both the extension on the interfacial plane ${\mathcal D}$ and the height $h$ of the microgel perpendicular to the interface (see Methods for the definition of these observables). 
\begin{figure}[t!]
\centering
\includegraphics[width=\textwidth]{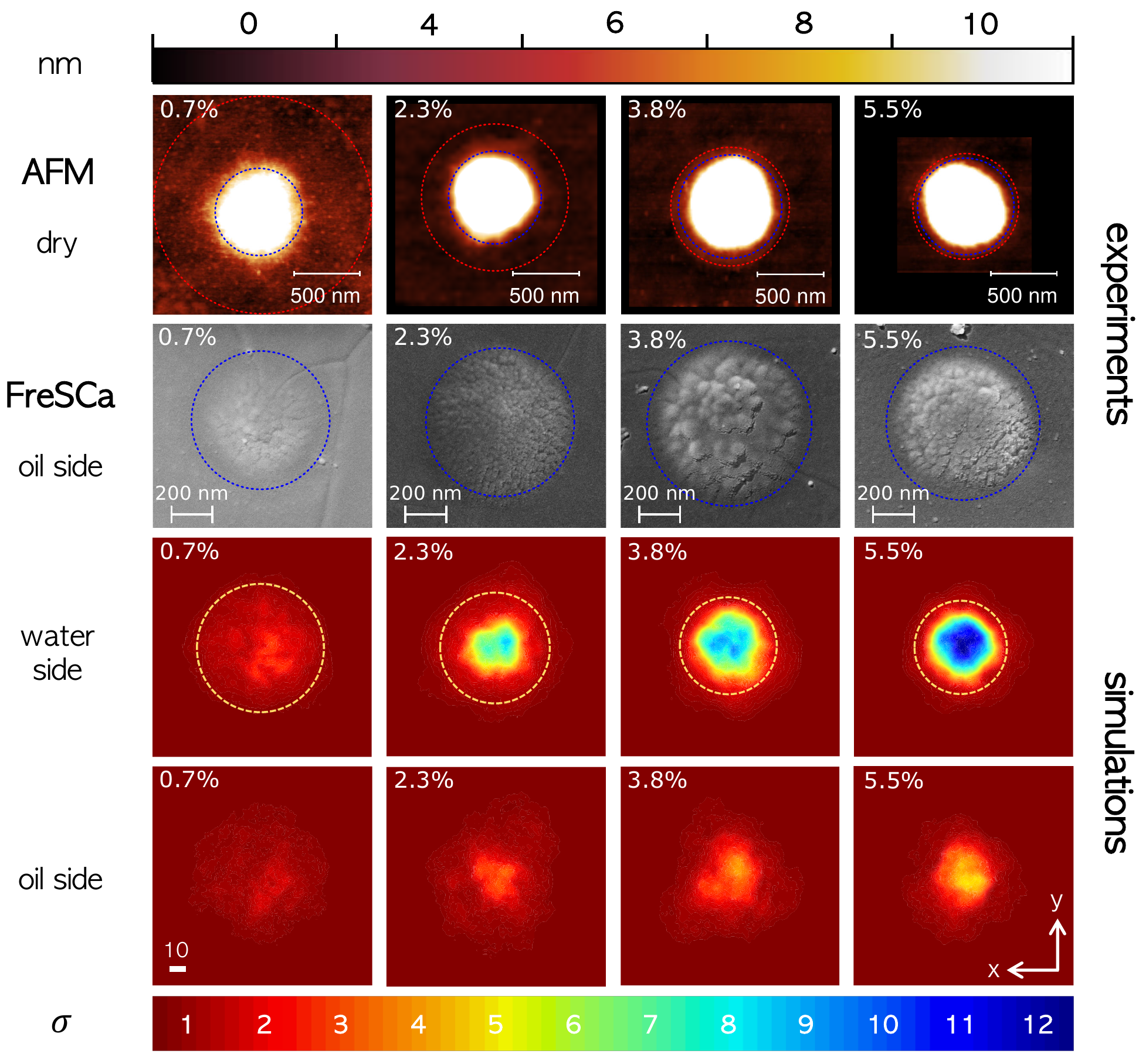}
\caption{\small \textbf{Conformation of microgels at the interface in experiments and simulations by varying $\textbf{c}$.} Top row: AFM height images of dried isolated microgels deposited from the water/hexane interface onto a silicon substrate. The top colour bar represents the height measured with AFM in $nm$. The height scale is saturated at 10 nm in order to clearly show both the thin corona and the higher core in the same image; second row from top: corresponding cryo-SEM images obtained by the FreSCa technique showing a frontal view of the interface with the microgels protruding into the oil phase, after removal of the latter. Red circles correspond to the average diameter measured from the AFM images, and blue circles from the FreSCa cryo-SEM images. It is evident that FreSCa cryo-SEM visualizes the core only; third and fourth rows from top: numerical surface plots of the microgels from the plane of the interface ($z=0$) towards the water and oil phases, respectively. Yellow circles are representative of the average extension taken for each of the cross-linker ratios analyzed. The bottom colour bar refers to the height of the numerical height profiles for both water and oil sides in units of $\sigma$.}
\label{fig:heightprof}
\end{figure}

Figure~\ref{fig:heightprof} shows AFM images of microgels with different $c$ after spreading at the water/hexane interface and deposition onto a silicon substrate. We also report FreSCa cryo-SEM images that provide a picture of the microgel upon vitrification of water and removal of the oil, as well as the numerical surface profiles from the water and oil sides of the interface. In FreSCa micrographs, the visible part of the microgel is the one protruding from the water phase into the oil. The outer corona is not visible with this technique as the low density of the dangling polymers makes it difficult to achieve sufficient contrast in the SEM imaging. Furthermore, since microgels do not cast any shadow (see experimental details in Methods) following tungsten coating at a $30^{\circ}$ angle, it can be seen that their effective contact angle is below $30^{\circ}$ and that they are mostly immersed in water~\cite{geisel2012unraveling}. Comparing the microgel size from the FreSCa cryo-SEM images (Table~\ref{tab:AFM_vs_FreSCa} and blue circles in Fig.~\ref{fig:heightprof}) and the AFM data (Table~\ref{tab:AFM_vs_FreSCa} and red circles in Fig.~\ref{fig:heightprof}), we see that the measured size $\mathcal{D}_{FreSCa}$ closely corresponds to the size of the more densely cross-linked core part of the microgel. Moreover, the data show that the thickness of the corona relative to the core size becomes smaller as the cross-linking ratio increases, which is expected since a more cross-linked microgel presents less dangling polymer chains.
\begin{table}[h!]
\footnotesize
\begin{tabular}{|c||c|c|c|}
\hline
\makecell{c} & \makecell{$\sigma_H$} & \makecell{$\mathcal{D}_{AFM}$} & \makecell{$\mathcal{D}_{FreSCa}$} \\\hline
0.7 & \makecell{628$\pm$165\\} & 1618$\pm$98 & 652$\pm$45 \\\hline
2.3 & \makecell{618$\pm$83\\} & 1095$\pm$66 & 700$\pm$62 \\\hline
3.8 & \makecell{597$\pm$127\\} & 882$\pm$30 & 777$\pm$39 \\\hline
5.5 & \makecell{574$\pm$73\\} & 786$\pm$30 & 724$\pm$29 \\\hline
\end{tabular}
\caption{\small \textbf{Experimental characterization in bulk and comparison between AFM and FreSCa.} Size of the microgels measured by DLS $\sigma_H$ and their extension after deposition on a silicon wafer measured by AFM $\mathcal{D}_{AFM}$, as well as their extension at the interface from FreSCa cryo-SEM measurements $\mathcal{D}_{FreSCa}$, for different values of the cross-linking ratio $c$ . Data are expressed in $nm$.}
\label{tab:AFM_vs_FreSCa}
\end{table} 
\begin{figure}[b!]
\centering
\includegraphics[width=\textwidth]{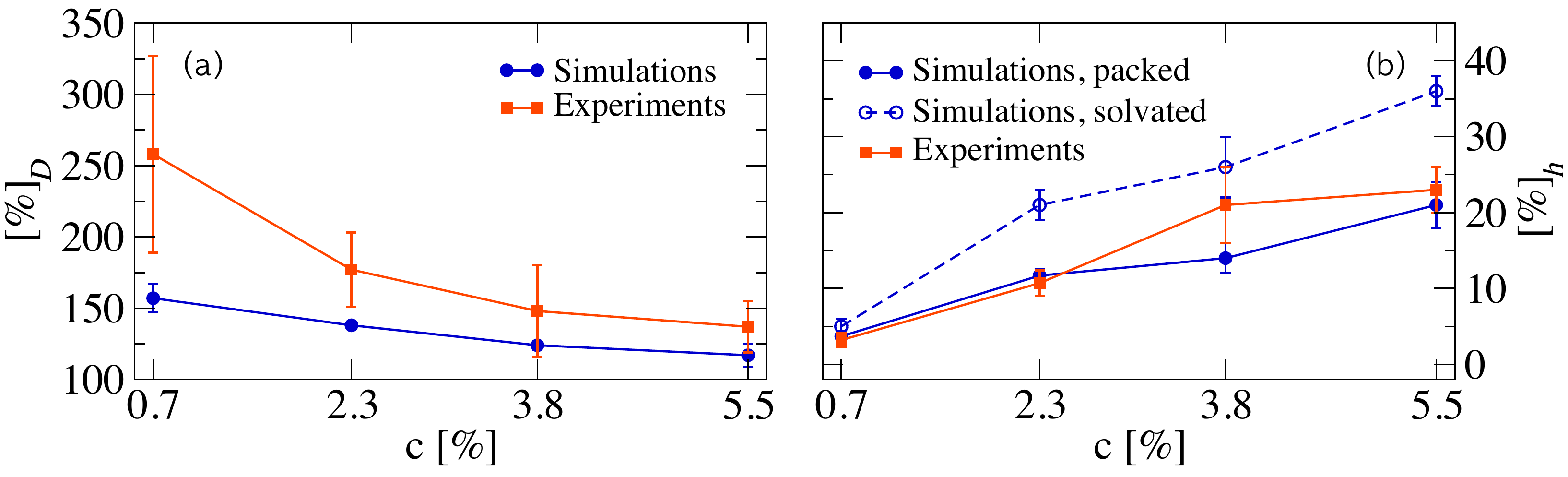}
\caption{\small \textbf{Comparison between simulations and experiments.} (a) Extension ratio $[\%]_\mathcal{D}$ with respect to the bulk diameter $\sigma_H$ for the cross-linker concentrations $c$ analyzed, for simulations (blue circles) and experiments (orange squares); (b) height ratio $[\%]_{h}$ with respect to the bulk diameter $\sigma_H$, measured by AFM for the dry microgel (orange squares) and calculated from simulations for the fully solvated microgel (empty blue circles) and for the packed microgel configuration (full blue circles).}
\label{fig:trends}
\end{figure}
\noindent
\\
A comparison between experimental and numerical results is provided in Fig.~\ref{fig:trends} where dimensionless observables are used. In particular, we define the following ratios: $[\%]_{\mathcal D}$ quantifies the increased extension of the microgel size at the interface with respect to its bulk value $\sigma_H$, while $[\%]_{h}$ represents the ratio of the microgel height with respect to $\sigma_H$. In the Supporting Information, a detailed table of all quantities presented here is reported.
\\
Starting with the analysis of the interfacial extension of the microgel, we find a qualitative agreement between experiments and simulations confirming that, by increasing the cross-linker concentration, microgels are less extended at the interface. Indeed, increasing $c$ the polymer network becomes stiffer due to the fact that polymer chains are closer to each other and less free to diffuse around. At the interface this translates into a more compact shape. At the same time, the more the corona contracts, the more the core of the microgel becomes denser and protrudes towards the water phase, as evidenced also in the density profiles and in the height profiles for both simulations and experiments.
An important contribution to the total extension of the microgel is given by the flattening of the corona at the interface. As it can be noticed by the height profiles in Fig.~\ref{fig:heightprof}, the spreading is responsible for the increase of $\approx 50-60\%$ of the total extension within the interfacial plane. At low cross-linker concentration, a true core can no longer be distinguished and this ratio certainly increases.

It is also important to note that the experimental height of the microgel grows by almost six times moving from $c=0.7\%$ to $c=5.5\%$, as shown in Fig.~\ref{fig:trends}(b). Interestingly, the cryo-SEM images in Fig. ~\ref{fig:heightprof} show an increase in height also in the oil side, where a visible protrusion, due to the core, is observed at high cross-linker concentration.
Since we cannot obtain the equivalent of the AFM dry height in simulations, we provide two different estimates of $[\%]_{h}$, reporting the calculation of the height for the fully solvated microgel and for the packed network configuration. While the former is obtained by taking into account solvated microgel configurations, the latter is estimated by projecting all the monomers down on the interfacial plane, stacking them onto each other (see Methods).
These two  quantities bracket the experimental results, showing a good agreement especially for the low cross-linked microgels. Moreover, they follow a similar trend as a function of cross-linker concentration.
\\
We notice that a systematically higher extension is found in experiments with respect to numerical results, which might be due to the way in which the size of the microgel is quantified in the two cases and/or due to size effects in simulations. Indeed, our {\it in silico} microgels are relatively small to correctly take into account the overall extent of the corona. Although we are able to maintain a realistic core-to-corona ratio in terms of their relative extension, we found a significant difference in the maximum chain length, particularly those of the corona or the so-called dangling ends\cite{ninarello2019advanced}. This may explain the smaller extension of the microgel at the interface with respect to experiments, where the outer dangling polymers are taken into account up to the limit of AFM resolution.
To further address this point, we tested larger -- yet far from experimental conditions -- microgels, observing very minor changes in the trends, despite a large increase in computational cost (see Supporting Information). Nonetheless, this additional study 
provides robustness to our approach, confirming the consistency of the method and the presence of a clearly identifiable ``fried-egg'' shape (Fig. S4), which is much more pronounced 
for the larger microgels.

\subsection{Effects of a different surface tension}

We now examine the results for a liquid pair with a different surface tension, to prove the soundness of our approach. In particular, we analyze the case of a water/benzene interface, whose measured surface tension\cite{zeppieri2001interfacial} is approximately 30\% lower than the one of water/hexane.
AFM results for microgels deposited on the water/benzene interface are reported in Table~\ref{tab:benzene}. The qualitative behavior of the microgel configuration is similar to that observed for water/hexane interface, with a decreasing extension and increasing height of the microgel for increasing cross-linking ratio.
Quantifying the difference between the two interfaces, we observe that for $c \gtrsim$ 3\% there is substantially no change of the microgel configuration within the experimental errors. A larger difference is observed for small $c$, particularly for the 0.7\% case, where the explored change in surface tension is able to modify the response of such a loose polymer network.
\begin{table}[b!]
\footnotesize
\begin{tabular}{|c||c|c|c|c|}
\hline
\multicolumn{5}{|c|}{\textbf{\makecell{ \\ Experiments\\ \\}}} \\
\hline
\makecell{c} &  \makecell{$\mathcal{D}$} & $[\%]_\mathcal{D}$  & $h$ & $[\%]_{h}$ \\
\hline
0.7 & 1474$\pm$62 & 234$\pm$62 & 22$\pm$2 & 3.5$\pm$1.0 \\
\hline
2.3 & 923$\pm$64 & 149$\pm$23 & 58$\pm$4 & 9.4$\pm$1.4\\
\hline
3.8 & 869$\pm$48 & 146$\pm$32 & 135$\pm$9 & 23$\pm$5\\
\hline
5.5  & 724$\pm$68 & 126$\pm$20 & 160$\pm$27 & 28$\pm$6\\
\hline
\end{tabular}
\caption{\small {\textbf{Experimental results at the water/benzene interface}} reporting the extension $\mathcal{D}$ and dry height $h$ of the microgel at the interface for $c=0.7, 2.3, 3.8, 5.5\%$. Data are given in $nm$. Ratios are in $\%$, referring to the bulk size $\sigma_H$.}
\label{tab:benzene}
\end{table}

\begin{figure}[h!]
\centering
\includegraphics[scale=0.45]{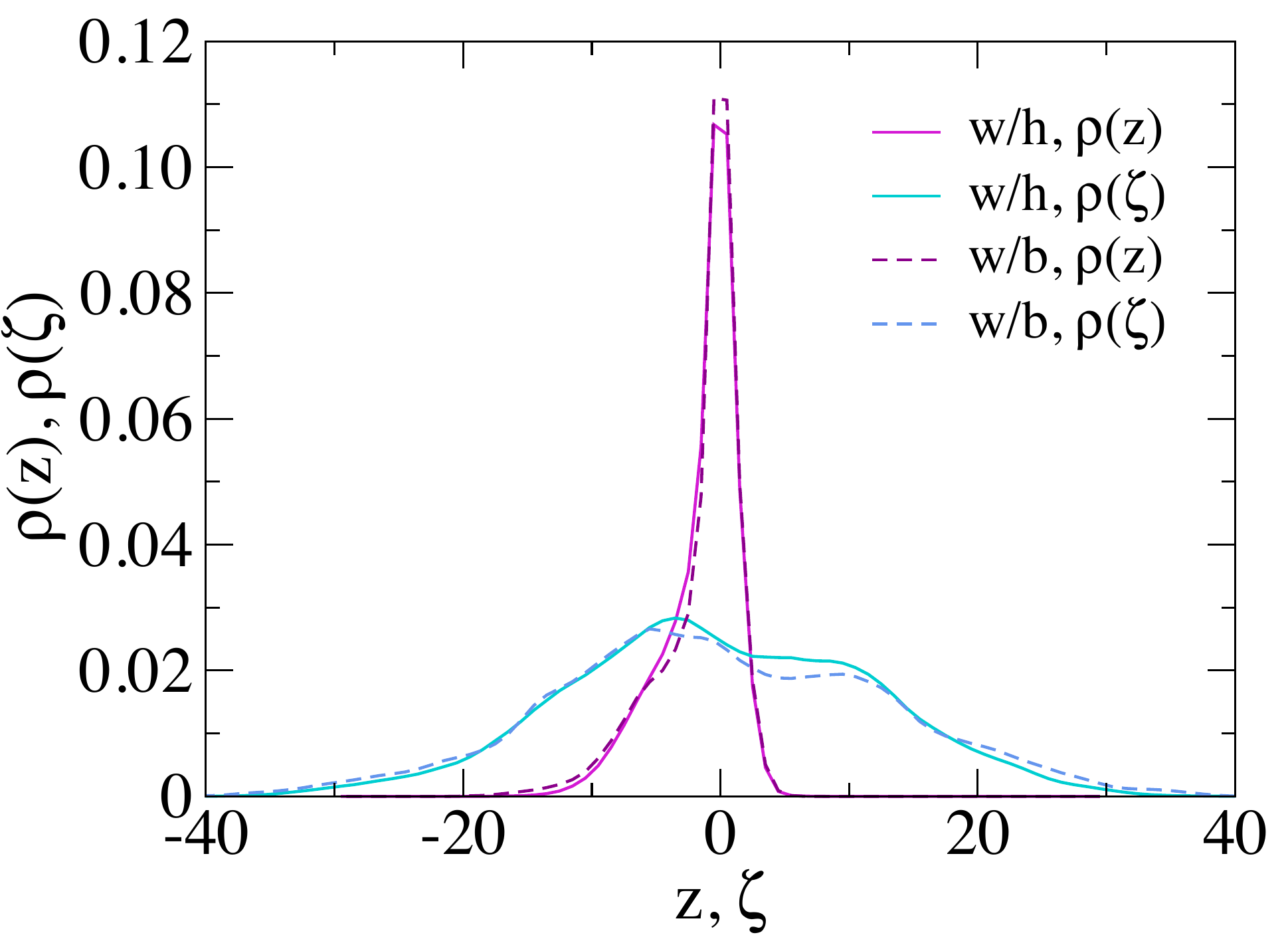}
\caption{\small \textbf{Comparison between a microgel at water/benzene and water/hexane interfaces.} Effects on $\rho(z)$ and $\rho(\zeta)$ of a different surface tension between different combinations of immiscible liquids, corresponding to water/hexane (w/h) and water/benzene (w/b), for $c=$3.8\%.}
\label{fig:hexbenz}
\end{figure}

To provide a comparison with simulations, we incorporate the effect of the surface tension between different combinations of fluids in the DPD modeling, as discussed in Methods.  Since it is reasonable to assume a similar solubility of PNIPAM in both hexane and benzene, the microgel-solvent interaction parameters have been correspondingly re-mapped (see Methods). As expected, the monomer-solvent interaction parameters have different absolute values due to the change in the surface tension, but they are still chosen in an intermediate range of the swelling curve in bulk (see Supporting Information).
Figure~\ref{fig:hexbenz} shows $\rho(z)$ and $\rho(\zeta)$ calculated for the same microgel configuration at the water/benzene and water/hexane interfaces for the cross-linker concentration $c=3.8$\%. Despite the reduction of the surface tension by roughly 30\%, we observe only very small differences in the distribution of monomers for the two interfaces. In particular, the extension on the interfacial plane does not vary significantly, in full agreement with the experimental results.
This holds especially for the most widely adopted fractions of cross-linkers $c\gtrsim3$\%, suggesting that, even for the lowest analyzed surface tension, an equilibrium between spreading and internal elasticity can still be reached.
Similar effects have been found when studying the microgel conformation at the interface as a function of pH\cite{geisel2012unraveling} and of temperature below the VPT\cite{monteux2010poly}. The microgel extension appears to be almost ``saturated'' in the studied cases, suggesting that we may only get significantly different results for much lower surface tensions or for large variations of the solution conditions. This point still deserves further investigation. 

\section{Conclusions}

The emerging potentialities of soft particles, particularly microgels, at liquid-liquid interfaces require microscopic models that reproduce the most relevant features observed experimentally.  In this work, we put forward an accurate theoretical description of a single microgel confined at the interface between two immiscible fluids.
We have first discussed the possible choices for water-monomer and oil-monomer interactions, discriminating cases where the microgel does not adsorb at the interface and those where the configuration is too compact. We interpreted these different scenarios as resulting from the balance between the adsorption and the elastic free energy contributions. We thus determined the optimal conditions under which the microgel maximizes its extension on the plane of the interface and protrudes toward the water phase. In this way, we have been able to reproduce the characteristic ``fried-egg'' shape at the interface and the increased flattening of the microgel when the number of cross-linkers decreases. Such behavior is not observed for regular networks such as the diamond one, being devoid of a well-defined core.
Moreover, we found that the numerical results are robust to size effects and are also valid for different values of the surface tension. Interestingly, the microgel configuration does not change significantly between a water/hexane and a water/benzene interface, despite a 30\% variation of the surface tension.
\\
Our modelling is consistent with experimental evidences on several aspects.
Firstly, we tuned the simulated surface tension to reproduce the one of a water/hydrocarbon interface by adjusting the repulsion parameters and the density of a coarse-grained DPD fluid. Such a solvent description may be exploited in other calculations of particles and polymers at interfaces, given its flexibility in the choice of the involved parameters. 
Moreover, our study built on a microgel model 
whose bulk swelling behavior, density profiles and form factors are directly comparable to the experimentally measured ones\cite{gnan2017silico,ninarello2019advanced}. Notably, the internal degrees of freedom are taken into account to reproduce the disordered polymeric network in a realistic way, also with respect to its inhomogeneous density profile, made of a denser core and a fluffier corona. By comparing numerical and experimental results, we found confirmation of the important role played by the corona in determining the extension of the microgel on the plane of the interface. 
\\
The aspects we have dealt with are particularly relevant in perspective.
From a fundamental point of view, the microscopic model presented here constitutes the basis for the numerical study of more complex assemblies. To investigate the physical origin that underlies the formation of two dimensional or quasi-two-dimensional structures, it is necessary to evaluate microgel-microgel effective interactions on the interface. Similar calculations were recently performed in bulk\cite{rovigatti2018validity,bergman2018new} and it will be interesting to compare the two cases. In the study of the macroscopic properties of microgel monolayers deposited at an interface, the role of the corona-corona interactions will be of fundamental importance. A further perspective  
will be the examination of the rheological properties of thin microgel monolayers for which the typical behavior of soft glassy materials has been 
lately observed\cite{huang2017structure,deshmukh2015hard}.
\\
Regarding applications, the emulsion-stabilizing effect as well as the use in surface patterning are only some of the recent advances that have been proposed\cite{thorne2011microgel,xia2017fabrication,tang2018surface}. For instance, patterned thermoresponsive polymer coatings have been identified as a valuable tool in bio-medicine for non-invasive control over cell-adhesion\cite{uhlig2018thermoresponsive}. For these and other purposes, the microscopic understanding of a two dimensional interfacial system, from single-particle studies up to collective behavior, is expected to strongly advance the field\cite{isa2017two,bresme2007nanoparticles} and is the basis for the development of nano- and micro-structured materials.

\section{Methods}
\small
\subsection{Simulations}
\noindent
\textbf{\textit{In silico} microgel synthesis.} Microgels are numerically synthesized in a fully-bonded, disordered network following the protocol described in Ref.~\cite{gnan2017silico}. The network is obtained by exploiting the self-assembly of $N_2$  and $N_4$ particles of diameter $\sigma$with two and four attractive patches, respectively, which mimic, in a coarse-grained fashion, the monomers (NiPAm, N-isopropylacrylamide) and cross-linkers (BIS, N,N-methylenebis(acrylamide)) of a PNIPAM microgel. The assembly process is carried out until more than $99.9\%$ of bonds are formed, so that all cross-linkers employed in the simulations are incorporated in the network. 
The polymeric nature of the network is enforced by adopting the Kremer-Grest bead-spring model~\cite{grest1986molecular}, whose non-bonded units interact \textit{via} a Weeks-Chandler-Andersen (WCA) potential and bonded ones \textit{via} a sum of the WCA and the Finitely Extensible Nonlinear Elastic (FENE) model, which are defined as:
\begin{equation}\label{wca}
V_{WCA}(r)=\begin{cases}
    4\epsilon\left[\left(\frac{\sigma}{r}\right)^{12}-\left(\frac{\sigma}{r}\right)^{6}\right] + \epsilon  & \text{if $r \le 2^{\frac{1}{6}}\sigma$}\\
    0 &  \text{otherwise}
  \end{cases};
\end{equation}
and
\begin{equation}
V_{FENE}(r)=-\epsilon k_FR_0^2\ln\left[1-\left(\frac{r}{R_0\sigma}\right)^2\right]     \text{ if $r < R_0\sigma$}
\end{equation}
where $\epsilon$ sets the energy scale, $k_F=15$ and $R_0=1.5$ the maximum extension that polymer bonds can reach.  For the present study, we use microgels with $N\approx 5000$ particles with unit mass $m$ confined within a sphere of diameter $Z=25 \sigma$, closely resembling the internal polymer density of real microgels. 
We study microgels with four different cross-linker concentrations $c$, that are $0.7, 2.3, 3.8$ and $5.5$\%, where $c$ is defined as $N_4$/$N_2$, in analogy with experiments. To keep a realistic core-corona relative extent, microgels are assembled employing an additional designing force that acts on the cross-linkers only~\cite{ninarello2019advanced}. In this way, we are able to reproduce a radial density profile that comprises of an inner core smoothly decreasing towards the periphery of the particle in close agreement to experiments. Size effects are assessed using microgels of $N\approx 42000$ with $Z=50 \sigma$, as discussed in the SI.
\\
We also perform tests with a microgel generated on regular diamond-based network with $N\approx 5000$ monomers and $c=5\%$. In this case the sites of the lattice represent the cross-linkers that are in turn connected by fixed-length polymer chains. By appropriately cutting the generated network, the microgel assumes a spherical shape.

\bigskip
\noindent
\textbf{Solvent modelling.} The numerical simulation of a liquid-liquid interface requires explicit solvent modeling in order to reproduce the surface tension effects between the two solvents. For this reason, we follow the same protocol described in Ref.~\cite{camerin2018modelling} by adopting Dissipative Particle Dynamics (DPD) simulations~\cite{groot1997dissipative}, which describe the solvent in a coarse-grained fashion. In brief, the DPD solvent is considered as made of soft beads interacting with each other \textit{via} a force that comprises conservative $\vec{F}^C_{ij}$, dissipative $\vec{F}^D_{ij}$ and random $\vec{F}^R_{ij}$ forces, which take the form
\begin{equation}\label{dpd_fc}
\vec{F}^C_{ij}=\begin{cases}
    a(1-r_{ij}/r_c)\hat{r}_{ij} & \text{if $r_{ij}<r_c$},\\
    0 & \text{otherwise}
  \end{cases}
\end{equation}
\begin{equation}
\vec{F}^D_{ij}=-\xi w^D(r_{ij})(\hat{r}_{ij} \cdot \vec{v}_{ij}) \hat{r}_{ij}
\end{equation}
\begin{equation}
\vec{F}^R_{ij}=\sigma_R w^R(r_{ij}) \theta (\Delta t)^{-1/2}\hat{r}_{ij}
\end{equation}
where $\vec{r}_{ij}=\vec{r}_i-\vec{r}_j$ with $\vec{r}_{i}$ the position of particle $i$, $r_{ij} = |\vec{r}_{ij}|$, $\hat{r}_{ij} = \vec{r}_{ij} / r_{ij}$, $r_c$ is the cutoff radius, $\vec{v}_{ij}=\vec{v}_i-\vec{v}_j$ with $\vec{v}_{i}$ the velocity of particle $i$, $a$ is the maximum repulsion between two particles, $\theta$ is a Gaussian random number with zero mean and unit variance and $\xi$ is the friction coefficient. To ensure that Boltzmann equilibrium is reached, $w^D(r_{ij})=[w^R(r_{ij})]^2$ and $\sigma^2_R=2\xi k_BT$ with $k_B$ the Boltzmann constant and $T$ the temperature. More details on the DPD implementation may be found in Ref.~\cite{groot1997dissipative}.
\\
Further exploiting the potentialities of DPD in treating mesoscopic systems, we refine this approach to mimic the experimental interfacial tension between the two solvents. To this aim, we adapt the work of Rezaei and Modarress\cite{Rezaei2015}, that focuses on finding the most suitable ``beading'' procedure for the two solvents as well as on correctly choosing the DPD parameters based on the Flory-Huggins mixing parameter $\chi$~\cite{rubinstein2003polymer}. We thus average the molecular volumes of the two liquids in order to have the smallest possible bead size. The resulting simulated fluid retains the features of both solvents. By estimating $\chi$, the DPD interaction parameters for the liquids are readily obtained. In order to mimic a water/hexane (w/h) system, we use $a_{\rm ww}=a_{\rm hh}=8.8$ and $a_{\rm hw}=31.1$ with a cutoff radius $r_c=1.9\sigma$ and a reduced solvent density $\rho_{\rm DPD}=4.5$. The surface tension $\gamma$ can be expressed in terms of the diagonal components of the pressure tensor as\cite{Maiti2004,Groot2001}
\begin{equation}
\gamma=\frac{1}{2}L_z \left[ p_{zz}-\frac{p_{xx}+p_{yy}}{2} \right]
\end{equation}
where $L_z$ is the measure of the side of the simulation box perpendicular to the interface; the $x$ and $y$ components define the plane of the interface. Under the chosen simulation conditions, we find $\gamma \approx 50$ mN/m in close agreement to the measured one\cite{zeppieri2001interfacial}.
\\
Equally important is the choice of the microgel-solvent interaction parameters that define the propensity of the microgel of being soluble. 
We note that, differently from standard DPD methods, microgel-monomers interaction parameters cannot be directly related to the Flory-Huggins mixing parameter, because we use the bead-spring potential to appropriately describe interactions among monomer beads. For this reason, the DPD parameters between monomers and solvent are chosen as discussed in the dedicated subsection of ``Results and Discussion''. We find that, by setting $a_{\rm mw}=4.5$ and $a_{\rm mh}=5.0$, the microgel equilibrates by protruding only in water and having the maximum extension on the plane of the interface.
\\
We also present results for simulations at the water/benzene (w/b) interface for which the DPD interaction parameters have to be adjusted. In this case, the reduced solvent density is $\rho_{\rm DPD}=3.0$ with $a_{\rm ww}=a_{\rm bb}=16.7$, $a_{\rm bw}=90.1$ and a cutoff radius $r_c=1.5\sigma$, yielding a surface tension $\approx 35$ mN/m. Given the similar solubility of PNIPAM both in hexane and benzene, we map the monomer-solvent interaction parameters from the water/hexane system to obtain the different values. This is done by rescaling the bulk swelling curves and by taking the corresponding $a_{\rm ms}$. In this way, we obtain $a_{\rm mw}=10.2$ and $a_{\rm mb}=11.0$. We underline that similar choices would lead to results very close to the ones presented.
\\
We perform simulations at the interface and in bulk. The latter are carried out keeping the same solvent features as in interfacial simulations, to be taken as a reference to determine the extent to which microgels deform at the interface. The monomer-solvent interaction parameter $a_{\rm ms}$ is set to $1.0$ for (w/h) and $5.0$ for (w/b), so that the microgel assumes its most swollen configuration, according to the swelling curve shown in Figure~\ref{fig:swelling} (see Supporting Information for the water/benzene case).
\\
All molecular dynamics simulations are run using the LAMMPS simulation package\cite{plimpton1995fast}. The equations of motion are integrated with a velocity-Verlet algorithm. The reduced temperature $T^*=k_BT/\epsilon$ is always set to $1.0$  \textit{via} the DPD thermostat (acting on the solvents only)~\cite{camerin2018modelling}. Length, mass, energy and time are  given in units of $\sigma$, $m$, $\epsilon$ and $\sqrt{m\sigma^2/\epsilon}$, respectively. DPD repulsion parameters $a$ are in units of $\epsilon/\sigma$. Finally, in bulk simulations, the center of mass of the microgel is fixed in the center of the simulation box whereas, at the interface, only fluctuations in the $z$ direction are allowed. Numerical measures are obtained by averaging over three independent microgel configurations for each of the cross-linker ratios analyzed.

\bigskip
\noindent
\subsection{Experiments}
\textbf{Microgel synthesis.} Our N-isopropylacrylamide (NiPAm, TCI 98.0$\%$) based microgels were synthesized by a conventional method of precipitation polymerization in water\cite{PNIPAM}. First, MilliQ water was heated to 80$\,^{\circ}\mathrm{C}$, NiPAm and the cross-linker N-N'-Methylenebisacrylamide (BIS, Fluka 99.0$\%$), were dissolved in four different moles of cross-linkers to moles of monomers ratios such that $c$=0.7, 2.3, 3.8, 5.5$\%$ with a total monomer concentration of 180 mM. For clarity, the related weight fractions are also provided in Table~\ref{tab:mols}. It is important to stress that each nominal ratio is only an upper limit of the actual incorporated amount of cross-linker in the final microgels, because usually some residual material is left over during the polymerization reaction.
Our monomer was purified and recrystallized in a solution of toluene/hexane, in a volume/ratio 60/40. The mixture was degassed with $N_2$, and the reaction started after the addition of potassium persulfate (KPS, Sigma-Aldrich 99.0$\%$) at 1.8 mM. The mixture was kept under stirring at 80$\,^{\circ}\mathrm{C}$ for 5h and, after that, it was let to cool down to room temperature. In order to clean the microgel dispersion, it was centrifuged three times at 20000 rpm for 1h, replacing the supernatant with MilliQ water after each centrifugation step and placing the dispersion under ultrasonication for 1h to redisperse the particles after each supernatant replacement, reaching a final concentration of $1  wt\%$. The size of the synthesized microgels was characterized in bulk MilliQ-water by dynamic light scattering (DLS, Zetasizer, Malvern UK) at 25$\,^{\circ}\mathrm{C}$ and the dependence on the bulk size with temperature was studied by measuring their size at different temperatures, from 25 to 40$\,^{\circ}\mathrm{C}$, for all the cross-linker concentrations (see Supporting Information).
\begin{table}[h]
\centering
\footnotesize
\begin{tabular}{|c|c|c|c|}
\hline
\multicolumn{1}{|l|}{Cross-linkers (g)} & \multicolumn{1}{l|}{Monomer (g)} & \multicolumn{1}{l|}{wt \%} & \multicolumn{1}{l|}{moles of cross-linkers per moles of monomer, $c$ (\%)} \\ \hline
1                                      & 99                               & 1                          & 0.7                                                         \\ \hline
3                                      & 97                               & 3                          & 2.3                                                         \\ \hline
5                                      & 95                               & 5                          & 3.8                                                         \\ \hline
7                                      & 93                               & 7                          & 5.5                                                         \\ \hline
\end{tabular}
\caption{\small \textbf{Experimental ratio of cross-linkers as weight and mols ratios.}}
\label{tab:mols}
\end{table}

\bigskip
\noindent
\textbf{Deposition of the microgels from the water/hexane interface and AFM imaging.} The process to deposit the microgels from the MilliQ water/hexane (Sigma-Aldrich, HPLC grade 95$\%$) interface is described in a previous work~\cite{rey2016isostructural}. Silicon wafers (Siltronix, <100> 100 mm single polished side) were cut with a diamond pen into 2 cm x 2 cm pieces and placed under ultrasonication for 15 min in subsequent baths of toluene (Fluka Analytical, 99.7$\%$), isopropanol (IPA, Fisher Chemical, 99.97$\%$) and MilliQ water, drying them afterwards with $N_2$. Next, the silicon substrates were placed on the dipper arm and immersed under water in a KSV5000 Langmuir oil/water trough setup with a deposition well. The substrates formed a $30^{\circ}$  angle with the water/hexane interface. Next, the microgel dispersion was diluted in a 1:7:2 microgel dispersion:MilliQ water:IPA mixture and it was injected directly at the water/hexane interface using  a Hamilton microsyringe. The surface pressure was simultaneously measured with a platinum Wilhelmy plate and the microgel injection was stopped at values below 1mN/m to ensure the presence of  isolated microgels. Next, the silicon substrate was lifted slowly to deposit the microgels from the water/hexane interface onto the substrate. The isolated microgels deposited on silicon wafers were characterized by atomic force microscopy (AFM, Brucker Icon Dimension), in tapping mode, using cantilevers with 300 kHz resonance frequency and 26 mN/m spring constant, taking images of $20\times 20$ and $3\times 3$ $\mu m^2$, all of them at 512 px/line and scanning speed of 1-1.5 Hz. The in-plane size of the particles after deposition was obtained from the AFM height images by limiting the height range to 10 nm and manually fitting the microgel perimeter with a circle including as much as possible of the outer corona. The edge of the microgel is defined as the point where measure height is larger than the background noise, in the range $1-3$ nm. The diameters and height were extracted from 30 different microgels.

\bigskip
\noindent
\textbf{Freeze-fracture shadow-casting cryo-SEM.} Additionally, the microgels were characterized by freeze-fracture shadow-casting cryo-SEM (FreSCa) technique\cite{Fresca}. In this technique, microgel-laden interfaces are prepared by pipetting 0.5 $\mu l$ of a microgel dispersion at 0.1 wt$\%$ in MilliQ water in a customized copper holder and then covering it with 3 $\mu l$ of purified decane (Sigma-Aldrich, 99$\%$) to form the water/oil interface. The FreSCa cryo-SEM measurements are performed at the water/decane interface, but no significant differences are expected with hexane, due to similar values of interfacial tension and PNIPAM solubility. After closing the copper holder the microgel-laden interface is vitrified with a propane jet freezer (Bal-Tec/Leica JFD 030) and fractured in high-vacuum and cryogenic conditions. The fracture preferentially exposes the water/oil interface, which is coated by a 2 nm layer of tungsten with a $30^{\circ}$ angle relative to the interface (Bal-Tec/Leica VCT010 and Bal-Tec/Leica BAF060). Finally, the interface is imaged in a cryo-SEM (Zeiss Leo 1530). The size or extension at the water/decane interface was measured from 30 different microgels.

\bigskip
\noindent
\subsection{Estimate of the microgel size and height at the interface} 
To compare the numerical results with experiments, we analyze the microgel spreading by measuring its extension $\mathcal{D}$ on the $xy$-plane of the interface and its height $h$ in the $z$ direction. We take as a reference the corresponding microgel size in bulk, for which we measure the hydrodynamic radius $R_H$, that is experimentally determined by DLS, implying $\sigma_H=2R_H$. In simulations,  $R_H$ is not readily available and thus we adopt an operative definition. Namely, we consider the radial density profile $\rho_{bulk}(r)$ of the microgel, and take $R_H$ at the distance where $\rho_{bulk}(r)=10^{-2}$. The obtained values of  $R_H$ are roughly proportional to those that are obtained by using the gyration radius $R_g$ (easily computed in simulations as $\left( \frac{1}{N}\sum_i^N \left(\vec{r}_i -\vec{r}_{CM} \right)^2 \right)^{1/2} $ with $\vec{r}_i$ the position of the $i$-th monomer and $\vec{r}_{CM}$ the position of the center of mass of microgel), since it was experimentally observed that for microgels $\frac{R_g}{R_H}\approx 0.6$~\cite{senff2000influence,senff1999temperature}. 
 \\
As discussed above, quantitative measures of the particle size at the interface are carried out with an AFM after deposition on a silicon wafer. In this way, we obtain the maximum extension of the particle under the assumption that it matches the one after deposition, as previously established for microgels\cite{scheidegger2017compression}. From the AFM images, after solvent removal, we can also extract the particle height $h$, that corresponds to the projected polymer density profile onto the plane of the interface. Unfortunately, our experiments do not make it possible to access the solvated conformation of the microgel at the interface and its protrusion in either of the two liquids, a precious information that is accessible from the numerical simulations only.  However, we notice that, even under dry conditions, a small amount of water, up to about $10\%$, may still be retained in the polymer network.
\\
To best reproduce the experimental acquisition techniques, we numerically calculate the extension of the flattened particle at the interface ${\mathcal D}$ as the average maximum distance that opposite edges of the polymer reach on the $xy$-plane. For the height, we report two estimates, for the fully solvated microgel and for the packed network configuration. The first is computed by drawing a surface profile on the oil and water sides of the microgel; the sum of the distances from the plane of the interface defines the height of the microgel. The latter is instead obtained by accumulating on the plane of the interface the number of monomers that occupy a certain $(x,y)$ coordinate, independently of $z$. The above measures are provided with an error bar that accounts for the differences in the number of monomers and topology of the configurations over which we average.

\normalsize
\section{Acknowledgements}
F.C., L.R., N.G., A.N. and E.Z. acknowledge support from the European Research Council (ERC Consolidator Grant 681597, MIMIC). M.A.F.R. and L.I. acknowledge the financial support of the Swiss National Science Foundation Grant PP00P2 172913/1. 

\renewcommand{\thefigure}{S\arabic{figure}}
\setcounter{figure}{0}

\section{Supporting Information}

We provide separately a Supporting Information file containing the microgel characterization  \textit{via} DLS measurements in bulk aqueous conditions, the bulk swelling curve of a microgel in a fluid that retains the features of both water and benzene, additional results using a diamond-lattice-based microgel, a table reporting a detailed comparison between simulations and experiments and the main outcomes for a larger microgel at a water/hexane interface.

\bibliography{mybib2}



\section*{Supplementary material}

# Microgels Adsorbed at Liquid-Liquid Interfaces: A Joint Numerical and Experimental Study

## Supporting Information

Fabrizio Camerin,*,†,‡ Miguel Ángel Fernández-Rodríguez,¶ Lorenzo Rovigatti,§,†
Maria-Nefeli Antonopoulou,¶ Nicoletta Gnan,†,§ Andrea Ninarello,†,§ Lucio Isa,*,¶
and Emanuela Zaccarelli*,†,§

†*CNR Institute for Complex Systems, Uos Sapienza, Piazzale Aldo Moro 2, 00185 Roma, Italy*

‡*Department of Basic and Applied Sciences for Engineering, Sapienza University of Rome, via Antonio Scarpa 14, 00161 Roma, Italy*

¶*Laboratory for Interfaces, Soft matter and Assembly, Department of Materials, ETH Zürich, Vladimir-Prelog-Weg 5, 8093 Zürich, Switzerland*

§*Department of Physics, Sapienza University of Rome, Piazzale Aldo Moro 2, 00185 Roma, Italy*

E-mail: fabrizio.camerin@uniroma1.it; lucio.isa@mat.ethz.ch; emanuela.zaccarelli@cnr.it

**Microgel characterization in bulk aqueous conditions**

Experimentally, the size of the microgels in aqueous bulk conditions was characterized by Dynamic Light Scattering (DLS) and is reported in Fig. S1. The average microgel size

1

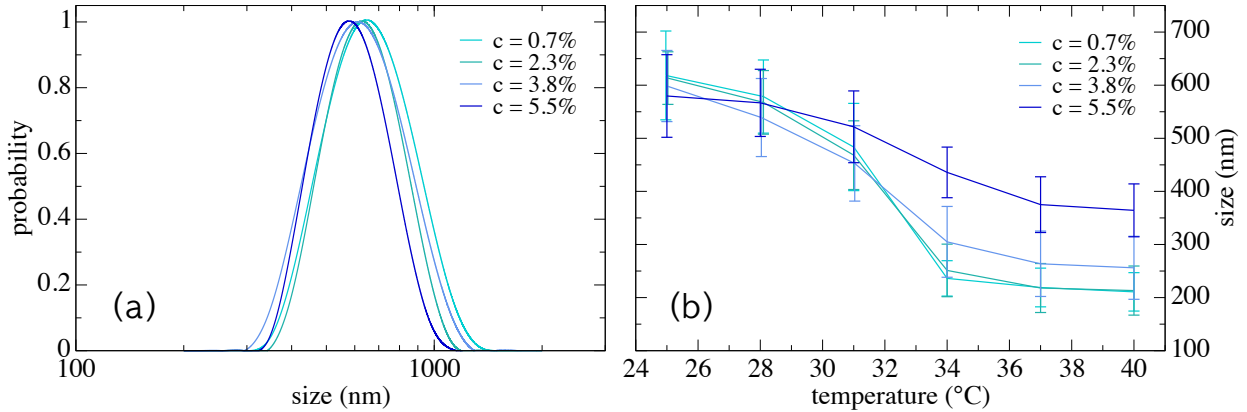

Figure S1: **DLS size characterization.** (a) Size distribution for the experimental microgels. (b) Size dependence on the temperature: as the cross-linking ratio increases, the microgels shrink less upon heating.

decreases with increasing cross-linking ratio. This is expected as more cross-linkers will produce a stiffer microgel that swells less in water. Moreover, this behaviour is further enhanced when looking at changes in microgel size with temperature. All microgels show a decrease of size above the characteristic volume phase transition temperature, $\sim 32\,°C$.[1] The size reduction is less pronounced for stiffer microgels, which shrink less than softer ones.

## Bulk swelling curve for the water/benzene case

We report in Fig. S2 the swelling curve of a single microgel in bulk, where the solvent has the same features of the interfacial water/benzene fluid.

## Diamond-lattice-based microgel

We perform simulations of a diamond-lattice-based microgel, generated as described in Methods, at the water/hexane interface. We consider $N \approx 5000$ monomers and $c = 5\%$. Snapshots of the microgel from different perspectives, to be directly compared with Fig. 1 of the manuscript for the disordered network, are provided in Fig. S3. The regular network characterizing this microgel is evident by looking at Fig. S3(a). A more detailed analysis of the internal structure as compared to the disordered polymer network can be found in Refs.[2–4]

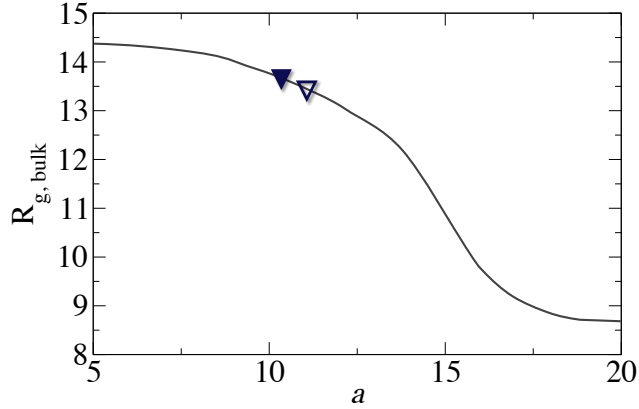

Figure S2: **Choice of the monomer-solvent interactions for the water/benzene interface**. Radius of gyration $R_{g,bulk}$ of a microgel with cross-linker concentration $c = 3.8\%$ in a one-component fluid with solvophobic parameter $a$ mimicking water/benzene interfacial fluid. The solvophobic parameters for the microgels at the interface are shown as triangles, with $a_{\rm mw} = 10.2$ for monomer-water interactions and $a_{\rm mb} = 11.0$ for monomer-benzene ones.

At the interface, the microgel spreads more with respect to the disordered configuration generated with our method, since the diamond network lacks a well-defined core. The increased number of monomers in the center of the microgel, that is visible by the side-view snapshots, is simply dictated by the accumulation of monomers towards the plane of the interface. This is also shown by the radial density profile reported in Fig. 3 in the main text. For these reasons, diamond-lattice-based models do not resemble the typical "fried-egg" shape observed experimentally in such systems.

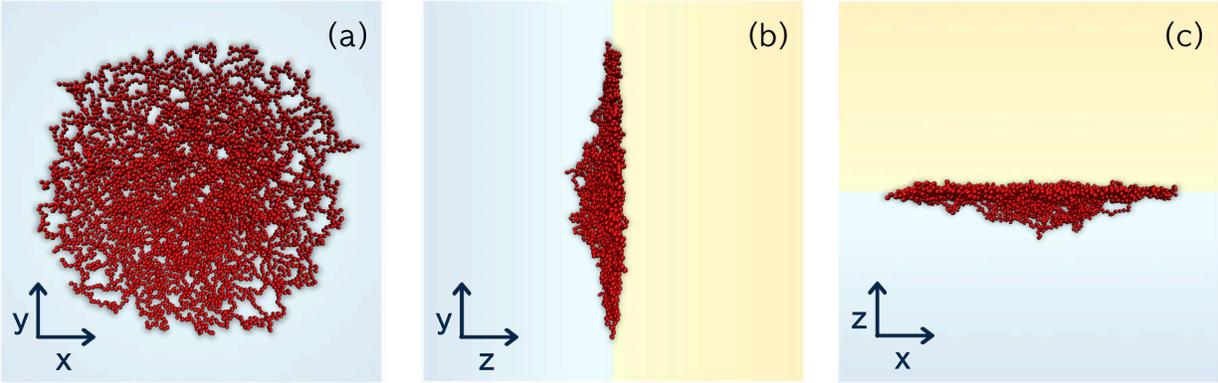

Figure S3: **Diamond based microgel network**. Simulation snapshots of a microgel built on regular diamond network with $N \approx 5000$ monomers and $c = 5\%$ at a water/hexane interface in the three different planes of observation: (a) top view (interface plane $xy$), (b,c) side views in which $z < 0$ corresponds to the water region and $z > 0$ to the oil one, respectively.

## Comparison simulations and experiments

We show in Table 1 a detailed comparison between simulations and experiments, as reported graphically in Fig. 6 in the main text. There, we also provide a definition of the quantities reported here.

Table 1: **Comparison simulations and experiments.** Summary table reporting: for experiments, the size of the microgels $\sigma_H$, the extension $\mathcal{D}$ and dry height $h$ of the microgel at the interface for $c = 0.7, 2.3, 3.8, 5.5\%$; for simulations, the solvated and packed heights $h$. Data are given in $nm$ for the experiments and in units of $\sigma$ for the simulations. Ratios are in %, always referring to the bulk size $\sigma_H$, which is taken from the bulk radial density profiles at $\rho = 10^{-2}$ for simulations and from DLS measurements for experiments.

| | Simulations | | | | | | | Experiments | | | | |
|---|---|---|---|---|---|---|---|---|---|---|---|---|
| c | $\sigma_H$ | $\mathcal{D}$ | $[\%]_\mathcal{D}$ | $h$ (packed) | $[\%]_h$ (packed) | $h$ (solvated) | $[\%]_h$ (solvated) | $\sigma_H$ | $\mathcal{D}$ | $[\%]_\mathcal{D}$ | $h$ | $[\%]_h$ |
| 0.7 | 53.6±0.3 | 85±5 | 158±10 | 2.0±0.2 | 3.7±0.4 | 2.9±0.5 | 5±1 | 628±165 | 1618±98 | 258±69 | 20±1 | 3.2±0.9 |
| 2.3 | 48.1±0.1 | 66.2±0.6 | 138±1 | 5.6±0.4 | 11.7±0.9 | 10±1 | 21±2 | 618±83 | 1095±66 | 177±26 | 66±6 | 10.7±1.7 |
| 3.8 | 45.8±0.2 | 57±1 | 124±3 | 6.4±0.9 | 14±2 | 12±2 | 26±4 | 597±127 | 882±30 | 148±32 | 125±12 | 21±5 |
| 5.5 | 43±1 | 51±2 | 118±8 | 9±1 | 21±3 | 15.4±0.7 | 36±2 | 574±73 | 786±30 | 137±18 | 133±7 | 23±3 |

## Microgel size effects

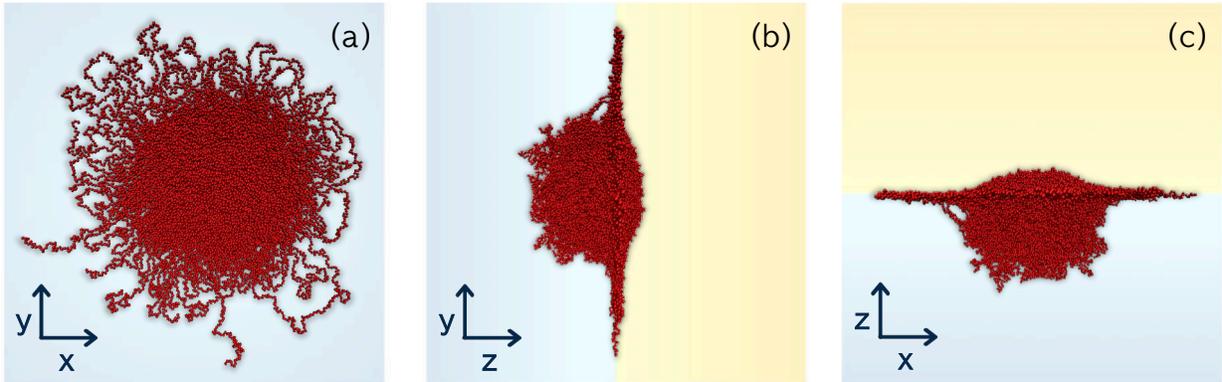

Figure S4: **A large microgel at a liquid-liquid interface.** Simulation snapshots of a large microgel with $N = 42000$ monomers and $c = 5\%$ at a water/hexane interface in the three different planes of observation: (a) top view (interface plane $xy$), (b,c) side views in which $z < 0$ corresponds to the water region and $z > 0$ to the oil one, respectively.

To test the robustness of the model presented in the main text we have also simulated a larger system with microgels made of 42000 monomers at a water/hexane interface with

$c = 3\%$ and $c = 5\%$. The microgel is assembled in the presence of an additional designing force to ensure the correct density profiles in bulk, following Ref.[5] Snapshots of the 5% microgel at the interface are shown in Fig. S4 from different perspectives, in full analogy with Fig. 1 of the manuscript. The way the microgel builds up at the interface is very similar to the smaller one, as evidenced by the $\rho(z)$ and $\rho(\zeta)$ (with $\zeta = x, y$) density profiles reported in Fig. S5. Also, the presence of a denser core and a thinner corona is confirmed. We observe a more evident core structure in the larger microgels, as signalled by the larger protrusion in $\rho(z)$. However, it is important to note that the interfacial extension does not vary significantly between the two structures. Indeed, we estimate $[\%]_\mathcal{D}$ to be 140% for $c = 3\%$ and 121% for $c = 5\%$. These values are in agreement with the trends reported for the smaller microgels in Table 2 of the manuscript. In addition, they are taken as described in the Methods section for smaller microgels and they are averaged over three independent configurations. However, we stress that in order to evaluate these properties for a $N \sim 42000$ monomers microgel in between the two solvents takes about 10 times in terms of computing time. Since the simulated is still very far away from the size of a real microgel and given the similarity of the results, in the manuscript we concentrate on the small microgel case.

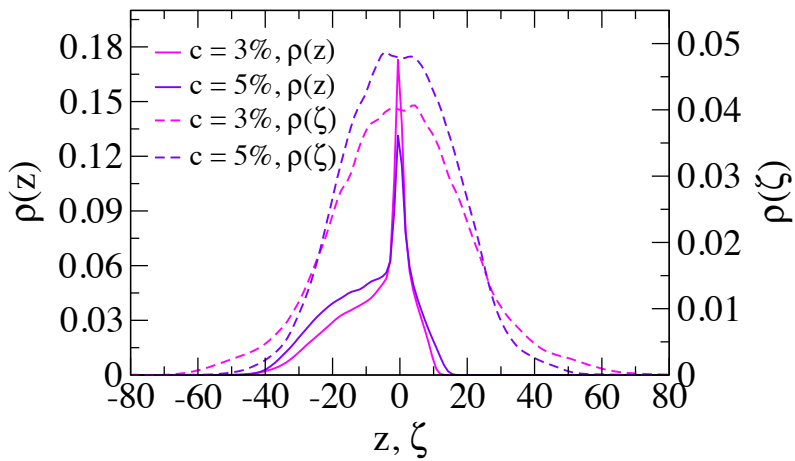

Figure S5: **Density profiles of larger microgels.** We show $\rho(z)$ and $\rho(\zeta)$ density profiles for microgels with $N \sim 42000$ microgel with $c = 3\%$ and $c = 5\%$.



\end{document}